\documentstyle{mn2e}
\input epsf
\def\plotone#1{\centering \leavevmode
\epsfxsize=\columnwidth \epsfbox{#1}}

\newcommand{\beq}{\begin{equation}}
\newcommand{\eeq}{\end{equation}}
\newcommand{\beqa}{\begin{eqnarray}}
\newcommand{\eeqa}{\end{eqnarray}}

\def\apj{ApJ}                 
\def\apjl{ApJ}

\def\deg{$^{\circ}$}

\title[Stability of cold fronts in clusters]{Stability of cold fronts
in clusters: is magnetic field necessary?}

\author[Churazov \& Inogamov]{E.~Churazov,$^{1,2}$ N.~Inogamov $^{3,1}$\\
$^1$ Max-Planck-Institut f\"ur Astrophysik, Karl-Schwarzschild-Strasse 1, 85741
Garching, Germany\\
$^2$ Space Research Institute (IKI), Profsoyuznaya 84/32, Moscow 117810, 
Russia\\
$^3$ Landau Institute for Theoretical Physics, Kosygin Street 2, V-334, 117940 Moscow,  Russia
}

\pagerange{\pageref{firstpage}--\pageref{lastpage}}
\pubyear{2003}

\begin{document}
\maketitle

\label{firstpage}
\begin{abstract}
Cold fronts -- sharp discontinuities recently discovered by Chandra in
many clusters of galaxies -- are believed to be due to a hot gas flow
over a colder gravitationally bound gas cloud. We analyze the
stability of the fronts with respect to Kelvin-Helmholtz instability
and show that an intrinsic width of the interface of the order of a few per
cent of the curvature radius strongly limits the growth of
perturbation. For the best studied case of a front in the Cluster Abell 3667 we
conclude that current observational data on the width and extent of
the front can be explained even in the absence of dynamically
important magnetic fields.
\end{abstract}


%

\sloppypar

\section{Introduction}
Cold fronts were discovered as sharp features in the X-ray surface
brightness distribution in Chandra observations of the clusters A2142
and A3667 (Markevitch et al., 2000, Vikhlinin, Markevitch \& Murray,
2001a), see also Markevitch et al. (2002). Similar features have now
been found in several other clusters (e.g. Sun et al., 2002, Kempner,
Sarazin \& Ricker, 2002). Unlike shocks, these features have lower gas
temperature on the X-ray brighter side of the discontinuity. For that
reason they are called ``cold fronts''. It is believed that some cold
fronts are formed when a subcluster merges with another cluster and
the ram pressure of gas flowing outside the subcluster gives the
contact discontinuity the characteristic curved shape. Indeed, features
resembling cold fronts are found in numerical simulations of cluster
formation (Bialek, Evrard \& Mohr, 2002, Nagai \& Kravtsov, 2003).
Ablation of the gaseous cloud by the hot gas causes characteristic
differential motion of the gas inside the subcluster, which transports
the low entropy gas from the subcluster core towards the contact
discontinuity, thus enhancing the jump in temperature and surface
brightness across the discontinuity (Heinz et al. 2003).

Here we address the question of the front stability. As was pointed
out by Vikhlinin et al. (2001a,b), Vikhlinin and Markevitch (2002) the
observed fronts are narrow and could be unstable to the
Kelvin-Helmholtz (KH) instability. For A3667 however the front appears
to be narrow (less than $\sim$5 kpc in width) up to $\sim$30\deg from
the stagnation point.  Magnetic fields can act as a stabilizing agent
thus allowing indirect estimates of the field strength near the front
(Vikhlinin, Markevitch \& Murray,2001b). On the other hand numerical
simulations without magnetic field (and in particular relatively high
resolution simulations by Heinz et al. 2003) do not show instability
of the front within 20-30\deg. While this discrepancy could be due to
numerical effects, we revisit the question of the front stability
below and show that the characteristic convex geometry and finite (small)
intrinsic width of the interface may help to stabilize the
discontinuity with respect to KH instability.

The structure of the paper is as follows. In Section 2 we derive
a simple expression for the growth of the KH instability along the curved
interface. In Section 3 we argue that diffusion processes are likely
to set an approximately constant width for the interface. In Section 4
we discuss astrophysical applications. The last section summarizes our
findings.

\section{KH instability along the spherical interface}
\begin{figure} 
\plotone{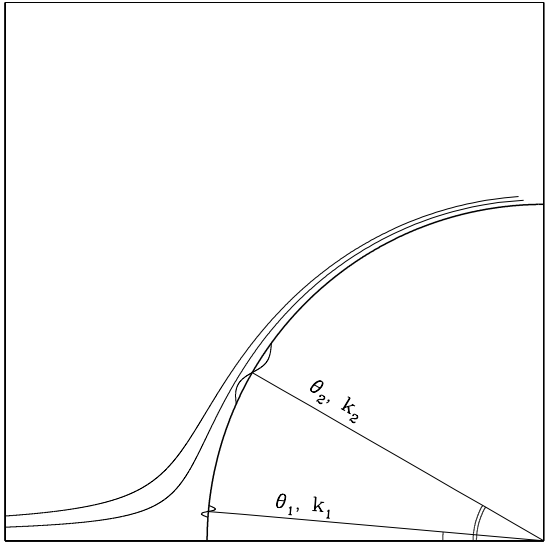}
\caption{Geometry of the problem. Hot gas is flowing around an
approximately spherical cloud of colder gas. Stream lines illustrate
the hot gas pattern of motion. The gas inside the cloud is assumed to
be still. While the interface is unstable against KHI, observations
shows that the front remains thin up to angles of
$\sim$30\deg. Perturbation with a wave number $k_1$ at angle
$\theta_1$ will have a wave number $k_2$ when arriving at angle
$\theta_2$ due to the increase of the  velocity along the
interface. 
\label{fig:geom}
}
\end{figure}

For an infinitely thin plane parallel interface separating
semi-infinite layers of two incompressible fluids with densities $\rho_1$
and $\rho_2$ the dispersion relation for Kelvin-Helmholtz instability
is (Landau \& Lifshitz, 1987):
\begin{eqnarray}
\omega=kv\frac{\mu\pm i \sqrt{\mu}}{1+\mu},
\label{eq:dr}
\end{eqnarray}
where $k=2\pi/\lambda$ is a wave number, $\lambda$ is the wavelength of
the perturbation, $v$ is the velocity of fluid with density $\rho_1$,
while fluid with density $\rho_2$ is at rest and
$\mu=\frac{\rho_1}{\rho_2}$. The frequency $\omega$ is measured in the
frame of the fluid at rest (i.e. with density $\rho_2$). For cold
fronts in clusters we adopt the geometry of the flow following
Vikhlinin et al. 2001a as shown in Fig.\ref{fig:geom}. The front has a
curvature radius $R$ and the velocity $v$ along the interface is
assumed to follow the law $v=v_0 \sin \theta$, where $\theta$ is the
angle along the interface. For the potential flow of incompressible fluid
$v_0=1.5~v_\infty$, where $v_\infty$ is the velocity of the cold gas
through the hot gas.  For a particular case of the cold front in A3667,
Vikhlinin et el. (2001b) argue that due to compressibility $v_0$ is
more close to $v_\infty$. For simplicity we assume that the densities of both fluids
remain constant along the interface.

Consider the perturbation of the interface separating cold and hot gas.
From eq.(\ref{eq:dr}) it follows that perturbations are advected along
the interface with the increasing velocity $v_a=v_{a0} \sin \theta$, where
$v_{a0}=v_0 
\frac{\mu}{1+\mu}$. This implies that the wave number $k$ of the perturbation
decreases as:
\begin{eqnarray}
k(\theta)=k_1 \frac{v_{a1}}{v_{a}}=k_1 \frac{\sin \theta_1}{\sin \theta},
\label{eq:stra}
\end{eqnarray}
where $k_1=k(\theta_1)$ is the wave number at some initial position
$\theta_1$. This result can be formally obtained from the lowest order
WKB (Wentzel-Kramer-Brillouin) approximation (see appendix),
applicable if $kR\sin\theta \gg 1$. 
Apart from the exponentially growing factor, the next order WKB
approximation allows one to estimate slower changes of the
perturbation amplitude $\delta$ caused by the changes of the velocity along the
interface and stretching of fluid elements in the azimuthal direction (see
appendix). For estimates we assume that $\delta \propto
\sin\theta^{-\alpha}$, and $\alpha\approx 2$. I.e. $\delta=\delta_1
\left ( \frac{\sin \theta_1}{\sin\theta} \right )^2$, where $\delta_1$ is 
the initial amplitude.

Thus one can evaluate the growth of the perturbation propagating along
the interface as:
\begin{eqnarray}
\delta=\delta_1  \left ( \frac{\sin \theta_1}{\sin
\theta} \right )^2 \exp \left \{ \int
\gamma(t)dt \right \},
\label{eq:growth}
\end{eqnarray}
where $\gamma(t)$ is the increment of KH instability and one has to
evaluate it using $k$ from eq.(\ref{eq:stra}). An analogous expression was
used by Inogamov \& Chekhlov (1991) for the growth of perturbations
for Rayleigh-Taylor instability.  From eq. (\ref{eq:dr}) and
(\ref{eq:stra}) it is clear that the increment $\gamma \propto kv$
does not increase as the perturbation propagates, since the increase
of the velocity amplitude is compensated by the decrease of the wave
number. Thus $\gamma(t)=k_1 v_{i0} \sin \theta_1=const$, where
$v_{i0}= v_{0}
\frac{\sqrt{\mu}}{1+\mu}$.
Replacing $dt$ with $dt=\frac{dt}{d\theta}d\theta=\frac{R}{v_a}d\theta$
one can write an explicit expression for $\int
\gamma(t)dt$:
\begin{eqnarray}
\int \gamma(t)dt=\int k_1 v_{i0} \sin \theta_1 \frac{R}{v_{a0} \sin
\theta}d\theta=\nonumber \\ 
R k_1 \sin \theta_1 \frac{1}{\sqrt{\mu}}
\ln  \left [ \tan(\theta/2)
\right ] .
\end{eqnarray}
Thus the final expression for the growth factor for a perturbation starting
at $\theta_1$ with the wave number $k_1$ measured at position $\theta_2$
is:
\begin{eqnarray}
{\rm Growth~factor}=\nonumber \\
\left ( \frac{\sin \theta_1}{\sin
\theta_2} \right )^2 {\rm \exp} \left \{ R k_1 \sin \theta_1 \frac{1}{\sqrt{\mu}}
\ln \left [ \tan(\theta/2) \right ] |_{\theta_1}^{\theta_2}   \right \}.
\label{eq:gft}
\end{eqnarray}
One can compare this expression with eq. (10) from Vikhlinin \&
Markevitch (2002) - i) an additional factor appears in front of the
exponential and more importantly ii) the argument of the exponential is changed
due to the variation of the  velocity along the interface, thus {\bf
reducing the growth factor} for a fixed $k_1$ and $\theta_1$.
We note here that from the point of view of 
the degree of nonlinearity of a perturbation one can use the ratio  
of the amplitude to the wavelength: $\delta/\lambda=\delta
\frac{k}{2\pi}$. Therefore the
expression for the degree of nonlinearity of a perturbation will
contain an additional factor proportional to
$\frac{k_2}{k_1}=\frac{\sin\theta_1}{\sin\theta_2}$ in front of the
exponential.
 
Alternatively, since $k_1 \sin \theta_1=k_2 \sin \theta_2$, we may
regard the growth factor as a function of $k_2$ specified at the angle
$\theta_2$.  The formal condition for the growth factor (5) to diverge
for small $\theta_1$ is $R k_2 \sin\theta_2 /\sqrt{\mu} > 2$.  Thus in
spite of the {\bf suppression} of the growth factor for a fixed {\bf
initial} wave number, the growth factor can be infinitely large for a
fixed {\bf final} wave number. There are however several factors which
can limit the growth of instability. We argue below that a small, but
finite, thickness of the interface is the most important factor in the
conditions relevant for cold fronts in clusters.

\subsection{Finite thickness of the interface}
\begin{figure} 
\plotone{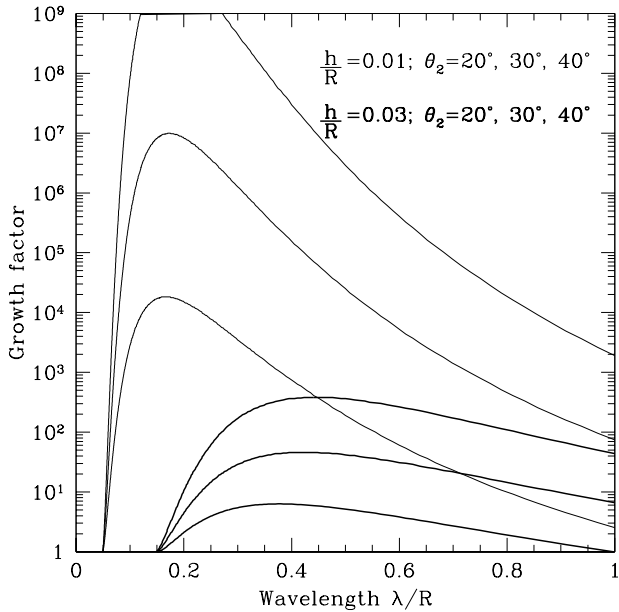}
\caption{Growth factor as a function of the wavelength $\lambda_2$ for
several values of angle $\theta_2$ and intrinsic with 
$h$. $\lambda$ and $h$ are expressed in units of the front curvature
radius. The factor $\frac{v_{i0}}{v_{a0}}$ in eq.(\ref{gfray}) is set
to 1.43.
\label{fig:growth}
}
\end{figure}

Any real interface can not be infinitely thin. If the thickness
of the interface is $h$, then only modes with $k \le k_{max}\sim
1/h$ are unstable. Thus for a fixed $\theta_2$ and $k_2<k_{max}$ the
minimum angle at which KH instability starts is such that $\sin
\theta_{min}=\frac{k_{2}}{k_{max}}\sin \theta_2$. In Section 3
we argue that diffusive processes across the interface may set a
finite thickness of the interface, which is almost independent of the
position (angle). One can approximately account for the finite width
of the interface by using $\theta_{min}$ as a lower bound in
eq.(\ref{eq:gft}). But in fact the largest contribution to the growth
factor comes from small angles and this part of the integral has to be
evaluated more accurately than simply introducing a cutoff in
$\theta$. One can use for instance the dependence of increment on the
wave number and thickness in the form given by Rayleigh (see "The
Theory of Sound", e.g. 1945 edition) for two fluids with similar
densities:
\begin{eqnarray}
\gamma(k,h)=\frac{v}{2h}\sqrt{e^{-2kh}-(kh-1)^2},
\label{ray}
\end{eqnarray}
where $k\le 1.2785/h$ is the condition for instability. Thus the growth
factor is:
\begin{eqnarray}
{\rm GF}=\left ( \frac{\sin \theta_1}{\sin
\theta_2} \right )^2
 {\rm \exp} \left \{\frac{R}{h} \frac{v_{i0}}{v_{a0}}
\int_{\theta_1}^{\theta_2}  
\sqrt{e^{-2kh}-(kh-1)^2} d\theta \right \},
\label{gfray}
\end{eqnarray}
where $\theta_1=\theta_{min}$ is such that $k_2\frac{\sin\theta_2}{\sin
\theta_1}=1.2785/h$. 
In this case the growth factor has to be calculated numerically. In
Fig.\ref{fig:growth} we show the total growth factor calculated for
several values of $\theta_2$ and $h$. For $\theta_2\sim$ 30\deg the
instability grows very strongly if the thickness of the
interface is less than $\sim$1.5-2\% of the curvature radius. 
We note that all the above
expressions for the growth factor do not depend explicitly on the
absolute value of $v_0$, indicating that the appearance of the front
may be similar for clouds moving with different velocities, if the
relation between the width and the curvature of the interface is the
same.

An interesting question is if the nonlinear evolution of KH instability
itself can be responsible for establishing the width of the interface
(along the lines of reasoning given by Nulsen, 1982) of order of few
percent of $R$, smoothing an interface and suppressing growth of longer
modes. We leave this question for subsequent studies and in the next
section consider the width of the interface set by diffusion.

\section{Width of the front due to transport processes}
For flow past a flat plate (with high Reynolds number) the width
of the boundary layer is known to vary as the square root of the
distance $x$ from the front edge of the plate (Blasius law):
\begin{eqnarray}
h\sim \sqrt{x \nu/U},
\end{eqnarray}
where $\nu$ is the viscosity.  For a rounded body, the flow velocity
increases linearly with distance from the stagnation point, $U\sim v_0
x/R$, making the width of the boundary layer constant (Landau \&
Lifshitz, 1987). Thus $h\sim \sqrt{\nu R/v_0}$, where $R$ is the
curvature radius. Substituting further $\nu\sim v_i \lambda_i$, where
$v_i$ and $\lambda_i$ are the rms velocity and mean free path for ions
and assuming that $v_i\sim v_0$ one gets an order of magnitude
estimate of the expected width $h\sim\sqrt{\lambda_i R}$. In the absence
of magnetic field the viscosity of the hot gas is larger than for
the cold gas and most of the velocity drop may take place on the
hotter side in the form of a broad layer. We note here that
observationally it is almost impossible to "see" such a layer.

Analogously one can consider the case of a scalar diffusion
(e.g. thermal conduction) with a constant diffusion coefficient
$D$. Assuming that conditions inside the cloud (where the fluid is not
moving) are frozen and seeking a stationary solution of a diffusion
equation in outer gas it is easy to show that $h \sim
\sqrt{\frac{DR}{v_{0}}}$.

Both expressions, when applied to the case of A3667 and compared with
the upper limit on the front width (Vikhlinin et al., 2001a), indicate
significant suppression of transport processes across the front. More
accurate calculations and a detailed comparison with observations
are beyond the scope of this paper and will be reported elsewhere. In
what follows we assume that the interface has a constant width
independent of $\theta$ and the expression (\ref{gfray}) for the
growth factor is applicable.
\section{Application to A3667}
\begin{figure} 
\plotone{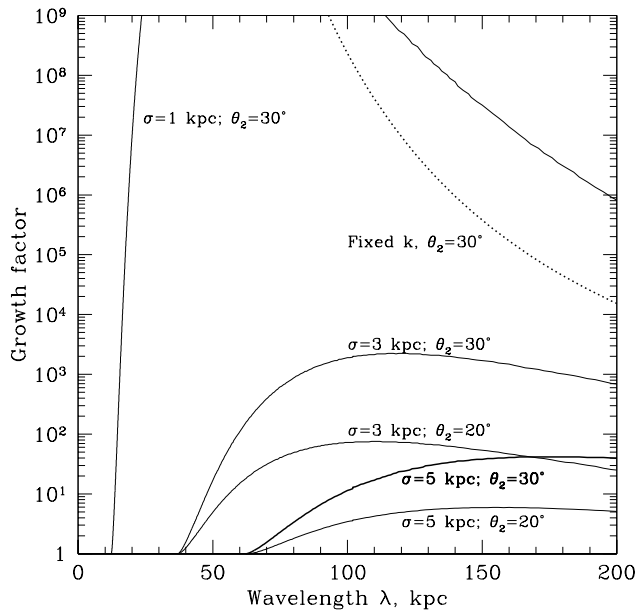}
\caption{Growth factor for A3667 parameters. Observational limit on
the width of the interface is $\sigma\le 5$kpc. Upper dotted curve
corresponds to the estimate of the growth rate for fixed $k$.
\label{fig:a3667}
}
\end{figure}
We now apply the above results to the particular case of the cold
front in A3667. We set the following parameters based on the results
of Vikhlinin et al., 2001a,b, Vikhlinin \& Markevitch 2002: $R=410$
kpc, $\theta_2=30$\deg, $\frac{v_{i0}}{v_{a0}}=1.43$. They give the
width of the front (upper limit) in terms of the standard deviation of
a gaussian smoothing, as $\sigma \le 5$ kpc.  In our notation,
$h=\sqrt{2\pi}\sigma\sim 2.5
\sigma$.  In Fig.\ref{fig:a3667} we plot the growth factor calculated
for several values of $\sigma$ and angle $\theta_2$. This is largely
the same figure as Fig.\ref{fig:growth} but now in natural units. For
comparison we show (upper dotted curve) the growth factor calculated
for an infinitely thin interface, assuming fixed wave number $k$ and
$\theta_2=$30\deg. One can see that accounting for varying wave number
and the finite thickness of the interface (within the observational
limits, i.e. $\sigma \le 5$ kpc) drastically reduces the growth
factor. For these reduced values it is not obvious that small initial
perturbations can grow strongly nonlinearly and disturb the sharp
appearance of the front around $\theta_2\sim$30\deg. For larger angles
or for smaller widths of the interface the growth factors are large.

\subsection{Role of the cloud internal motions}
While we assumed above that the gas inside the cloud is still,
internal motions are likely to appear during the formation of the front
and especially during later periods of the cloud evolution when
continuous stripping at the sides of the cloud is compensated by the
cold gas flow along the interface to replenish the losses.  E.g. in
simulations by Heinz et al. (2003) the velocity of the gas inside the
cloud $\sim$0.2 of the velocity of the surrounding gas at the moment
when the morphology of the front and the temperature distribution resemble
those observed in A3667\footnote{We note here that in the absence of
gravity the velocities of the colder and hotter fluids would be
related as $v_{cold}=v_{hot}\sqrt{\mu}$ as follows from the Bernoulli
equation. Due to gravity the velocity of the motions in cold gas
should be lower.}. The internal motions have a two-fold effect: i)
they reduce the shear rate and ii) they increase the group speed of
the perturbation in the interface, both of which reduce the total
growth of a perturbation arriving at the angle $\theta_2$. Accounting
for internal motions can easily reduce the argument of the exponential
in equation (\ref{gfray}) by 30-50\% thus strongly reducing the final
growth factor compared to Fig.\ref{fig:a3667}.

\subsection{Role of gravity and compressibility}
In the above discussion we have  completely neglected the role of
gravity, although we assumed that it plays an important role in
preserving the integrity of the cold cloud when it passes through the
hot gas. For an infinitely thin interface, gravity
modifies the dispersion relation (\ref{eq:dr}) to:
\begin{eqnarray}
\omega=kv\frac{\mu}{1+\mu}\pm k\sqrt{\frac{g}{k}\frac{1-\mu}{1+\mu}-v^2\frac{\mu}{(1+\mu)^2}},
\end{eqnarray}
where $g$ is gravitational acceleration. For a fixed $k_2$ and
$\theta_2$ and given that $k=k_2\sin\theta_2/\sin\theta_1$ and
$v=v_0\sin\theta$ the role of gravity is more important for small
angles $\theta$, effectively introducing a lower limit on
$\theta_1$. The growth factor calculated with the above dispersion
relation and using the values of $g$ and $v_0$ from Vikhlinin \&
Markevitch (2002) is very large, implying that gravity does not 
strongly suppress KH instability, in agreement with their conclusion.

For an interface of finite thickness, gravity makes the region
near the stagnation point stable for perturbations of any $k$. One
can estimate the Richardson number of the flow assuming that
density and velocity both vary over a layer of thickness
$h$:
\begin{eqnarray}
Ri \sim \frac{\frac{-g}{\rho} \frac{\partial\rho}{\partial z}}{\left (
\frac{\partial v }{\partial z}
\right )^2 } \approx \frac{\Delta \rho}{\rho}\frac{gh}{v^2}.
\label{ri}
\end{eqnarray}
The sufficient condition for stability is $Ri>0.25$ everywhere in the
flow. Again plugging in the numbers from Vikhlinin \& Markevitch
(2002) we find that for $\sigma=1,3,5$ kpc the front is stable up to
angles of $\sim$6,11,14\deg respectively. Simple estimates show that
account for this stable region would further reduce the growth factor
compared to Fig.\ref{fig:a3667}, although not dramatically. This is
expected given that gravity alone (for the thin interface case - see
above) does not strongly suppress KH instability. Accurate
calculations of the growth factor taking account of gravity and the
finite thickness of the interface require knowledge of the density and
velocity profiles of the interface and are beyond the scope of this
paper. We note only that stretching of the flow in the stable region near
the stagnation point insures that initial perturbations in
the unstable regions further downstream are small.

Since we are considering motion near the stagnation point the
velocities are small compared to the gas sound velocity (even for
colder gas) and compressibility does not play a significant
role. Further downstream the compressibility starts to be more important
and it reduces the growth factor of the perturbation (see dispersion
relation in Miles, 1958, Gerwin, 1968 or Nulsen
1982). E.g. for the density ratio $\mu=0.5$ the quantity
$\frac{\rm Im[\omega]}{\rm Re[\omega]}$ is equal to
$\frac{1}{\sqrt{\mu}}=1.41$ for small shear velocity and it drops to
$\sim 1.13$ when the shear velocity approaches the sound speed of the
colder gas\footnote{An argument of the exponential in
eq.(\ref{eq:gft}) is proportional to this factor}.

\section{Conclusions}
We have shown that accounting for wave number changes along the curved
interface and finite (but small) width of the interface significantly
reduces the KH instability growth factor. We also argue that if
diffusion sets the intrinsic width of the interface then this width
does not vary much along the interface (at least for small angles).

For a set of parameters relevant for the front in A3667 the growth
factor is not large enough to guarantee nonlinear growth at angles
$\sim$30\deg, provided that the intrinsic width of the front is of the
order of a few percents of the curvature radius (i.e. not much less
than the existing observational limit). Therefore dynamically
important magnetic fields may not be necessary to stabilize this
particular cold front. The growth factor (and therefore possible
limits on the magnetic fields) {\bf depends critically on the width of
the interface} and much more weakly on the angular extent over which
the front remains sharp. A factor of 2-3 improvement in the
observational constraints on the interface width is therefore
extremely important.

For a given width of the interface the value of the growth factor does
not depend strongly on the absolute value of the shear velocity. This
probably explains the ubiquity of fronts both in observations and
hydrodynamical simulations.

\section{Acknowledgements} 
We are grateful to Sebastian Heinz, Rashid Sunyaev and Alexey
Vikhlinin for numerous useful discussions. We would like to thank the
referee, Paul Nulsen, for a number of important and constructive
comments and suggestions.

\appendix
\section{WKB approximation for plane accelerating flows}
Consider a plane boundary between two incompressible, inviscid,
irrotational fluids with densities $\rho_1$ and $\rho_2$ moving in the
$x$ direction with velocities $u_1(x,y)$ and $u_2(x,y)$. The fluid '1'
is accelerating along the interface: $\frac{\partial
u_1}{\partial x}=u_1/x=const$. To match pressures at the boundary we
have to assume that the velocity of fluid '2' is $u_2=u_1
\sqrt{\rho_1/\rho_2}=u_1\sqrt{\mu}$ and is also accelerating along the
interface. We now consider small perturbations of the boundary
$\xi(x,t)$ and velocity potentials of two fluids $\phi_{1,2}(x,y,t)$. The governing equations are:
\begin{eqnarray}
\Delta \phi_{1,2}=0 \\  
\frac{\partial \phi_{1,2}}{\partial y}=\frac{\partial \xi}{\partial t} + u_{x,1,2}\frac{\partial \xi}{\partial x} - \xi \frac{\partial u_{y,1,2}}{\partial y} \\
\rho_1 \frac{\partial \phi_{1}}{\partial t} + \rho_1 u_{x,1}\frac{\partial \phi_{1}}{\partial x} = \rho_2 \frac{\partial \phi_{2}}{\partial t} + \rho_2 u_{x,2}\frac{\partial \phi_{2}}{\partial x},
\end{eqnarray} 
where $u_{x,1,2}$ and $u_{y,1,2}$ are the velocity components in $x$
and $y$ directions for two flows. The second equation is the kinematic
boundary condition and the third equation is the Bernoulli
equation. These equations are evaluated at the boundary $\xi$.

\sloppypar

We are seeking a solution of these equations in the form $\xi=b(x)e^{i
(S_t(t)+S_x(x))/\varepsilon}$ and $\phi_{1,2}=a_{1,2}(x,y)e^{i
(S_{t}(t)+S_{x}(x)+S_{y,1,2})/\varepsilon}$, which decays for
$y=\pm\infty$, where $a$,$b$, and all $S$ are slowly varying functions
and $\varepsilon$ is a small parameter. Using these expressions and
keeping only the terms of order $1/\varepsilon$ we recover a
dispersion relation, which we write in a "spatial" form, denoting
$\frac{\partial S_t}{\partial t}/\varepsilon=-\omega$, $\frac{\partial
S_x}{\partial x}/\varepsilon=k$:
\begin{eqnarray}
k=w\frac{\rho_1 u_{x,1}+ \rho_2 u_{x,2} \pm i
(u_{x,1}-u_{x,2})\sqrt{\rho_1\rho_2}}{\rho_1 u_{x,1}^2+ \rho_2
u_{x,2}^2}
\end{eqnarray}
Now collecting higher order terms and using the relations
$\frac{\partial a_{1,2}}{\partial y}=\pm\frac{1}{i}\frac{\partial
a_{1,2}}{\partial x}$ from the Laplace equation and $\frac{\partial
a_2}{\partial x}=\frac{\rho_1 u_{x,1}}{\rho_2 u_{x,2}}\frac{\partial
a_1}{\partial x}$ from the Bernoulli equation we get a transport equation
for spatial variations of the amplitude $b$:
\begin{eqnarray}
(\rho_1 u_{x,1}^2+\rho_2 u_{x,2}^2)\frac{d b}{d x} - (\rho_1
u_{x,1}\frac{\partial u_{y,1}}{\partial y} + \rho_2
u_{x,2}\frac{\partial u_{y,2}}{\partial y})b=0.
\end{eqnarray}
Thus
\begin{eqnarray}
\frac{d \ln b}{d x}=\frac{\rho_1 u_{x,1}\frac{\partial u_{y,1}}{\partial y} + \rho_2 u_{x,2}\frac{\partial u_{y,2}}{\partial y}}{\rho_1 u_{x,1}^2+\rho_2 u_{x,2}^2}
\label{tr}
\end{eqnarray}
Incompressibility implies that $\frac{\partial u_{y}}{\partial
y}=-\frac{\partial u_{x}}{\partial x}=-u_{x}/x$. The last equality is
due to assumed form of acceleration. Therefore
\begin{eqnarray}
b\propto x^{-1}, 
\end{eqnarray}
i.e. the amplitude decreases with the distance as $1/x$. If additional
uniform stretching in the 3rd dimension is present then one has to set
$\frac{\partial u_{y}}{\partial y}=-\frac{\partial u_{x}}{\partial
x}-\frac{\partial u_{z}}{\partial z}$. In the case of interest for us
(potential flow past a sphere) one can simply set $\frac{\partial
u_{y}}{\partial y}=-2u_{x}/x$. Substituting this into eq. (\ref{tr}) we get
\begin{eqnarray} 
b\propto x^{-2}.
\end{eqnarray}

\label{lastpage}
\end{document}